\documentclass[aps,twocolumn,showpacs,floats,superscriptaddress]{revtex4}
\usepackage{graphicx}
\usepackage{dcolumn}
\usepackage{bm}
\def\he4{$^4$He}
\def\Am3{\AA$^{-3}$}
\def\beq{\begin{equation}}
\def\eeq{\end{equation}}
\begin{document}

\author{M. Boninsegni}
\affiliation{Department of Physics, University of Alberta,
Edmonton, Alberta T6G 2J1}

\author{A.B. Kuklov}
\affiliation{Department of Engineering Science and Physics,
CUNY, Staten Island, NY 10314}

\author{L. Pollet}
\affiliation{Theoretische Physik, ETH Z\"urich, CH-8093 Z\"urich, Switzerland}

\author{N.V. Prokof'ev}
\affiliation{Department of Physics, University of Massachusetts,
Amherst, MA 01003, USA}
\affiliation{BEC-INFM, Dipartimento di Fisica,
Universita di Trento, Via Sommarive 14, I-38050 Povo, Italy}
\affiliation{Russian Research Center ``Kurchatov Institute'',
123182 Moscow, Russia}

\author{B.V. Svistunov}
\affiliation{Department of Physics, University of Massachusetts,
Amherst, MA 01003, USA}
\affiliation{Russian Research Center ``Kurchatov Institute'',
123182 Moscow, Russia}

\author{M. Troyer}
\affiliation{Theoretische Physik, ETH Z\"urich, CH-8093 Z\"urich, Switzerland}

\title{The fate of vacancy-induced supersolidity in \he4}

\date{\today}
\begin{abstract}
The supersolid state of matter, exhibiting non-dissipative flow in solids, has been elusive for thirty five years. The recent discovery of a non-classical moment of inertia in solid \he4 by Kim and Chan provided the first experimental evidence, although the interpretation in terms of supersolidity of the ideal crystal phase remains subject to debate. Using quantum Monte Carlo methods we investigate the long-standing question of vacancy-induced superflow and find that
vacancies in a \he4 crystal phase separate instead of forming a supersolid. On the other hand, 
non-equilibrium vacancies relaxing on defects of poly-crystalline
samples could provide an explanation for the experimental observations.
\end{abstract}

\maketitle

\begin{figure}[t]
\centerline{\includegraphics[angle=-90, scale=0.35]{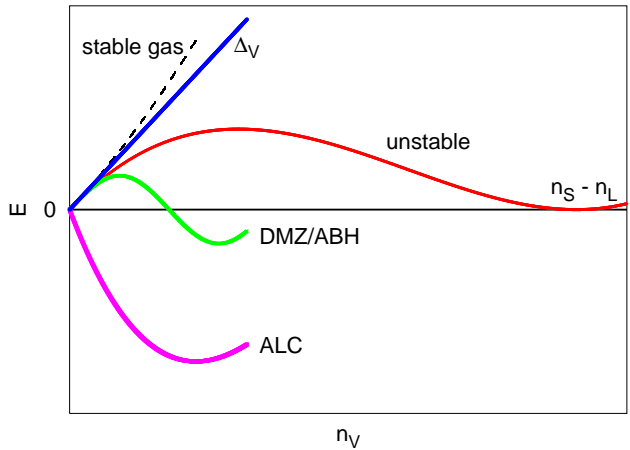}}
\caption{(Color online) A sketch of the different scenarios for vacancy states in the ground state of solid Helium at the melting density. Supersolidity can occur at finite vacancy densities $n_{\rm V}$, according to the Andreev-Lifshitz-Chester (ALC) or Dai-Ma-Zhang(DMZ)/Anderson-Brinkman-Huse (ABH) scenarios. ALC assume an energy gain already for a single vacancy (negative initial slope), while DMZ/ABH postulate a ground state with a low density of strongly correlated vacancies, without making any prediction for the single vacancy behavior. For the scenarios with gapped vacancies (positive slope $\Delta_{\rm V}$), a (meta)stable gas of out-of-equilibrium vacancies
will be formed if the energy curve bends up (the upper dashed line) due to repulsion between the vacancies. 
The curve labeled 'unstable' represents the situation supported in this paper: An attractive vacancy gas is unstable and phase separates into the commensurate
insulating crystal (with density $n_{\rm S}$) and the liquid
(with density $n_{\rm L}< n_{\rm S}$).}
\label{fig:scenario}
\end{figure}

The observation of a non-classical moment of inertia in
solid \he4 by Kim and Chan (KC) has provided the first experimental evidence of a possible supersolid phase of matter {  \cite{KCNature,KCScience}}, which is characterized by crystalline order and frictionless flow.
Early theories of supersolidity were based on the assumption that the
low-temperature \he4 crystal may be incommensurate (the number
of atoms is not an integer multiple of that of lattice sites).
As a consequence of their quantum behavior, point defects such as vacancies and interstitials
can Bose condense at low temperature, giving rise to superflow. 

In particular, the Andreev-Lifshitz-Chester (ALC)
scenario  \cite{andreev69,chester} assumes that the gain in kinetic energy by delocalizing the vacancy can overcome the potential energy cost of creating it in a perfect crystal, such that a dilute gas of highly mobile vacancies can be stabilized. The energetics of this scenario is illustrated in the lower curve in Fig.~\ref{fig:scenario}.
Similar supersolid ground states with a low density of strongly correlated vacancies were recently put forward by Dai, Ma and Zhang (DMZ)~\cite{Dai05} and also by Anderson, Brinkman and Huse (ABH) \cite{anderson}.
Their scenarios are possible even when single-vacancy excitations in the perfect crystal are
gapped, as indicated in Fig.~\ref{fig:scenario}. It was also recently suggested that the onset of supersolidity leads to anomalies in the elastic moduli~\cite{Dorsey06}.

Although the initial experimental observation by KC has been confirmed by other groups
  \cite{Rittner,kubota,shirahama}, the interpretation in terms of superflow in the crystal
phase is becoming increasingly questionable. Repeated cycles of annealing make the supersolid signal weaker to vanishing~\cite{Rittner}. Measurements of
pressure-driven flow have yielded a null result in
hexagonal-close-packed (hcp) \he4  \cite{Bi05,Bi06}. Experimental evidence  \cite{fraass,meisel} points toward
a commensurate hcp ground state in \he4, as the measured vacancy
activation energy is approximately $10-15$~K, and appears to rule
out thermally activated vacancies at the characteristic
temperatures below 0.2 K of the KC experiment. X-ray measurements \cite{fraass}
on hcp \he4 put the tightest upper bound on the vacancy
concentration $n_{\rm V}(T\to 0)$, disfavoring the ALC and DMZ/ABH
scenarios, although ABH debate the interpretation
of these measurements  \cite{anderson}. On the theoretical side, two of us have
proven that a superfluid crystal must generically be
incommensurate  \cite{theorem}, a view which is supported by recent numerical simulations
showing that ideal hcp crystals of \he4 are insulating  \cite{bernu,superglass,clark}.


The physics of vacancies in solid
Helium is an important open problem, even aside from its relevance to the experiment of KC, and has been studied for a long time. The majority of microscopic
calculations have been variational  \cite{pederiva97, pederiva99, galli}, and focused on the properties
of a single vacancy. Various estimates of the vacancy activation energy  \cite{pederiva97, pederiva99},
including the one obtained by Ceperley and Bernu using Path Integral Monte Carlo
simulations  \cite{bernu}, are in quantitative agreement with experiment.
However, Galli and Reatto questioned the reliability of all previous numerical
calculations, including their own ones  \cite{galli}, by raising the importance of finite size effects.

Here we present results of a Quantum Monte Carlo study of vacancies in hcp \he4 at low temperature (0.2 K $\le$ $T$ $\le$ 2 K).   For single crystals in the thermodynamic limit, we find large vacancy and interstitial activation energies $\Delta_{\rm V}$ and $\Delta_{\rm I}$ at all densities, ruling out the ALC scenario. We then proceed with the discussion on finite vacancy concentrations to see if activated vacancies form a (meta)stable gas (upper dashed line in Fig.~\ref{fig:scenario}), or if they support the DMZ/ABH-type state with a small vacancy concentration $n_{\rm V}$.  It turns out that the only minima in the grand-canonical (free) energy are at zero vacancy concentration, corresponding to a commensurate solid, and at high vacancy concentration such that the system is in a liquid phase. In Fig.~\ref{fig:scenario} this scenario is called ``unstable''.

%
We employ the
worm algorithm  \cite{worm}, a grand-canonical Path Integral Monte Carlo technique
formulated in the configuration space of the single-particle
Matsubara Green's function. In our simulations, we use a three-dimensional cell, designed to fit
a perfect finite crystal of $N$ \he4 atoms, with periodic boundary
conditions in all directions.
Our microscopic model of \he4 is
standard, based on the accepted Aziz pair potential  \cite{aziz}.
Single-particle properties can be deduced from the
Matsubara Green function
\begin{equation}
G({\bf r},{\bf r^\prime},\tau)\; = \; \langle \, {\cal T} \{
\hat \psi({\bf r},\tau)\, \hat{\psi}^\dagger({\bf r}^\prime,0) \}\, \rangle \;,
\label{GG}
\end{equation}
which can easily be sampled in the worm algorithm. Here
$\langle...\rangle$ stands for the thermal expectation value,
${\cal T}$ is the time-ordering operator, $-\beta /2 \le\tau \le
\beta /2$, $\hat{\psi}^\dagger ({\bf r}^\prime,0)$ and $\hat\psi({\bf r},\tau)$
are Bose particle creation and annihilation
operators, respectively. For $\tau > 0$ ($\tau < 0$)
one is computing the Green function for an interstitial atom (vacancy).

\begin{figure}[t]
\centerline{\includegraphics[angle=-90,scale=0.3]{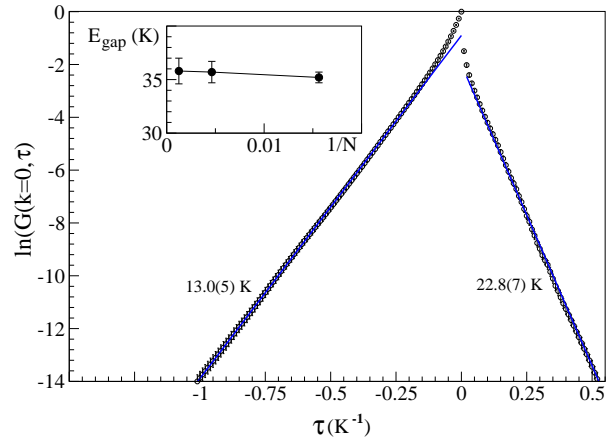}}
\vspace*{-0.5cm}
\caption{(Color online) Single-particle Green function $G({\bf k}=0,\tau)$
computed for hcp \he4 at the melting density $n_{\circ}$=0.0287 \Am3~
and $T=0.2$~K. Symbols refer to numerical data,
solid lines are fits to the long-time exponential decay.
The given numerical values are the interstitial ($\Delta_{\rm I}=22.8 \pm 0.7$ K)
and the vacancy ($\Delta_{\rm V}=13.0 \pm 0.5$ K) activation energies,
inferred from the slopes of $G$.
The inset shows the vacancy-interstitial gap $E_{\rm gap} = \Delta_{\rm I}+\Delta_{\rm V}$
 for different system sizes.}
\label{fig:green}
\end{figure}

In Fig.~\ref{fig:green} we show the spatially averaged Green function
$G({\bf k}=0,\tau)$ for the hcp \he4 crystal at melting
density $n=0.0287$ \Am3 and for a low temperature $T$=0.2 K. Results
are shown for a system of $N$=800 atoms. Since vacancies and interstitials are gapped by a finite activation energy, the low temperature Green function
$G({\bf k}=0,\tau)$ decays asymptotically as
\[
G({\bf k}=0,\tau) \sim e^{-|\tau| \Delta},
\]
with activation energies $\Delta =\Delta_{\rm V}$ ($\Delta_{\rm I}$) for $\tau < 0$ ($\tau >
0$), which are
measured relative to the chemical potential $\mu$ of the system. In order to
determine the chemical potential at the melting-freezing curve,
$\mu_{\circ}$, we have computed by simulation the equilibrium density of the liquid,
$n(\mu)$,  at low $T$. We located the experimentally known freezing point $n(\mu_{\circ} )=0.02599$
\Am3 with high accuracy.
Our estimate is $\mu_{\circ}=0.06 \pm
0.04$ K. Upon fitting the observed exponential decays of
$G(0,\tau)$, we obtain $\Delta_{\rm V}$=13.0(5) K, and $\Delta_{\rm I}$=22.8(7) K.
The value of $\Delta_{\rm V}$ found here is only slightly lower than
results from other numerical calculations, whereas that of $\Delta_{\rm I}$
is a factor of two smaller than the one reported in the only comparable
study  \cite{bernu}.

We did not find any appreciable size dependence of either $\Delta_{\rm V}$
or $\Delta_{\rm I}$, as seen in the inset of Fig.~\ref{fig:green}.
Since the results are also temperature independent for $T < 1.5$ K,
our data yield reliable ground state estimates in the thermodynamic limit. 
We can thus confirm that single vacancies and interstitials in a perfect
\he4 crystal have huge activation barriers at the low temperatures of the KC experiment. This finding is
consistent with the absence of thermally activated vacancies at low $T$, as reported in X-ray experiments \cite{fraass}.

\begin{figure}[t]
\centerline{\includegraphics[angle=-90,scale=0.3]{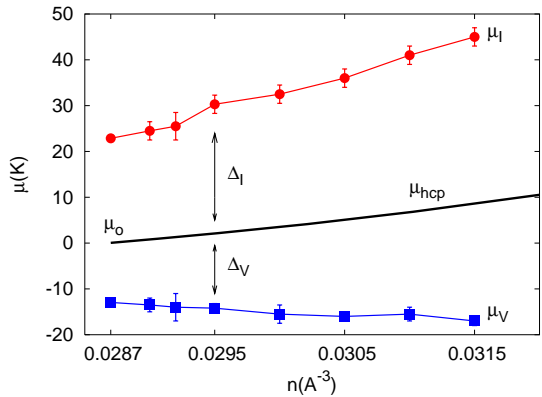}}
\caption{(Color online) Chemical potential thresholds for creating a single
interstitial (upper curve, $\mu_{\rm I}$) and vacancy (lower
curve, $\mu_{\rm V}$) at different densities. The middle curve,
$\mu_{\rm hcp}$ is the chemical potential of the hcp solid deduced
from the compressibility data  \cite{TOLYAKNOWS} and the value of
$\mu_{\rm hcp}=\mu_{\circ}$ at melting density. The gaps
$\Delta_{\rm V(I)}$ for activating a vacancy (or interstitial) are
given by the difference in chemical potentials, $\mu_{\rm
hcp}-\mu_{\rm V}$ (or $\mu_{\rm I} - \mu_{\rm hcp}$).}
\label{fig:threshold}
\end{figure}
The same Green function analysis was extended to higher densities.
From Fig.~\ref{fig:threshold} we infer that both $\Delta_{\rm V}$ and $\Delta_{\rm I}$ increase with
density, or equivalently, pressure.
This can be understood intuitively, as the cost for creating a defect
in a denser system is higher. Large activation energies thus prevent the ALC scenario in solid Helium at any density. It seems reasonable to extend this conclusion to all
presently known, naturally occurring solids since they have stronger
interatomic potentials.

We proceed to the physics of multi-vacancy states to check whether the highly correlated DMZ/ABH ground state in Fig.~\ref{fig:scenario} can be formed, despite the large activation energy $\Delta_{\rm V}$.
We
also explore the possibility of a (meta)stable gas of
out-of-equilibrium vacancies. The thermodynamic stability of a
uniform dilute gas of vacancies can be addressed by performing
simulations in the grand-canonical $(\mu-V-T)$ ensemble. We  have performed
simulations at low temperature $T$ = 0.2 K with $N$ = 800 and 2016 atoms in the crystal under different pressures up to 65 bars.  We manually created vacancies by explicitly removing a number
of atoms. Various initial configurations with randomly located, remote and clustered vacancy positions were considered, with vacancy concentrations ranging from half a percent up to six percent.

Our consistent observation is the formation of a commensurate hcp crystal,
as the grand-canonical simulation fills the vacancy sites with atoms.
This behavior is seen not only when the chemical potential $\mu$ is
set equal to the experimental chemical potential $\mu = \mu_{\rm hcp}$, but also when it 
is shifted several K below $\mu_{\rm hcp}$, which can be explained
by the large activation energy $\Delta_{\rm V}$. When the chemical
potential $\mu$ is lowered below that of
the liquid but is still kept above the single-vacancy threshold (
$\mu_{\rm V} < \mu < \mu_{\circ}$), a runaway in vacancy number
for all finite initial concentrations occurs, leading to a liquid
phase. 

We conclude that a uniform dilute gas of vacancies is thermodynamically
unstable against separation into a vacancy-rich 
and a perfectly crystalline 
vacancy-free phase, which does not melt:  The scenario corresponding to the curve labeled ``unstable'' in Fig.~\ref{fig:scenario} is realized in \he4, no DMZ/ABH-type state is found. The instability occurs not only at a finite vacancy concentration, but also for a few vacancies in a large crystal.

In order to confirm this conclusion, we performed simulations in the canonical $(N-V-T)$ ensemble, keeping the number of particles fixed. The instability translates into phase separation, with the vacancies forming clusters inside a perfect crystal. Indeed, our simulations produce the spontaneous formation of vacancy clusters for temperatures up to 1 K, regardless of pressure and the initial configuration. Phase separation is also supported by energy calculations, since the energy difference $(E(n_{\rm V}) - E(0))/n_{\rm V}$ is not monotonically increasing with vacancy concentration $n_{\rm V}$.

\begin{figure}[t]
\centerline{\includegraphics[angle=-90,scale=0.3]{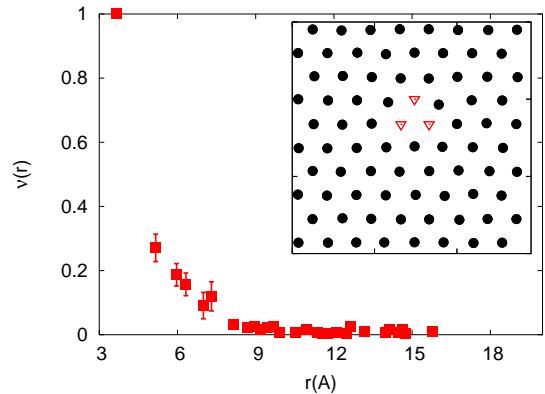}}
\caption{(Color online) The vacancy-vacancy correlation function $\nu(r)$ as a function of the distance $r$ between the vacancies shows that three vacancies easily cluster and form a tight bound state. The inset shows a typical snapshot of a layer of atomic positions averaged over the time interval $[0, \beta]$(filled black circles), where we see that the three vacancies (triangles) have a tendency to cluster in layers. }
\label{fig:vv}
\end{figure}

Inspection of the simulations revealed that already three vacancies cluster easily and form a tight bound state, as shown in the inset of Fig.~\ref{fig:vv}. The vacancy-vacancy correlation function is the quantity most sensitive to the sign of their effective interaction. We have computed it by first averaging atomic world lines over an imaginary-time scale of the order of 0.2 ${\rm K}^{-1}$, which is much longer than the zero-point atomic motion but much shorter than the vacancy-atom exchange time. Next, we compare the obtained atomic positions with the lattice points in an hcp crystal. Using a similar procedure as in Ref.  \cite{theorem}, we repeatedly remove the closest particle-lattice point pairs from the list, until we end up with a small number of unmatched lattice points which we define as the vacancy positions. The decay of the vacancy-vacancy correlation function $\nu(r)$ in Fig.~\ref{fig:vv} shows that the vacancy-vacancy interaction is attractive and that three vacancies form a tight bound state. 
A similar study
for two vacancies also yielded attractive correlations but was less conclusive regarding a bound state; if it exists, it is rather shallow with a binding energy of about 1K.

The observed attractive interaction between vacancies -- leading to phase separation -- arises from the elastic interactions. These are long ranged, decaying as $1/r^3$, and are attractive in certain directions (in the basal plane and along the $c$-axis) due to dominant quadrupolar terms  \cite{eshelby56,andreev76}.
A similar effect was previously suggested for bound states of substitutional $^{3}$He atoms \cite{richards75,mullin75}, and it is also reminiscent of a proposal for oxygen vacancies in high-temperature superconductors~\cite{khomskii01}. The strong effective short-range attraction is due to the local minimum in the pair potential.


Given the instability of the vacancy gas, the only remaining possibilities
accounting for the experimental data of KC are inhomogeneous
scenarios, due to a less than ideal sample quality  \cite{burovski,superglass}.
Metastable vacancies can play a pivotal role in this case.
When a large vacancy cluster is formed, the system can macroscopically lower
its energy by deforming the cluster into a dislocation loop. We found, by simulating
artificially created edge dislocations in the hcp crystal, that these
can be either insulating or sustain superflow depending on the annealing procedure for the initial
setup. In particular, in the presence of a large stress field
close to the dislocation core, the core first melts locally into a liquid,
which then forms a highly disordered superfluid (an undercooled superglass  \cite{superglass}).

Other scenarios include the migration of vacancies towards the grain boundaries
and boundary edges, along which superflow might be possible:
imagine, as a schematic example, an insulating grain boundary, where the potential produced by the crystal bulk forces the atoms at the interface to form an insulating Mott state. When vacancies relax on the grain boundary, they effectively dope the Mott insulator and may make it superfluid --- a phenomenon well documented in lattice models.

In conclusion, we have shown that vacancies are unstable in a \he4 crystal. They form
clusters and the system phase separates into a vacancy-rich phase
and a perfect, insulating crystal.
Recent experimental results~\cite{Rittner} are in perfect agreement with the fact that the ground state of solid \he4  is thus never a supersolid, but a commensurate crystal. The experimental observation of a non-classical moment of inertia in solid Helium is due to non-equilibrium quantum behavior, such as superflow along defects or the formation of a superglass around crystallographic defects. High resolution structural analysis of defects in the \he4 samples and large scale simulations of elementary defects will be important in completing our understanding.

We are grateful to P. W. Anderson, V. S. Boyko, W. F.~Brinkman, D. M. Ceperley, R. A. Guyer,  R. B. Hallock, D. A.~Huse, A. E. Meyerovich, W. J. Mullin,  and T. M.~Rice for stimulating discussions. This work was supported by the National Aero and Space
Administration grant NAG3-2870, the National Science Foundation
under Grants Nos. PHY-0426881, PHY-0456261 and PHY-0426814, the Natural Science and Engineering Research Council of Canada, under  research grant G121210893, the Swiss National Science Foundation, the Kavli Institute for Theoretical Physics in Santa Barbara and the Aspen Center for Physics. Parts of the simulations were performed on Beowulf clusters: Hreidar  at ETH Zurich and Typhon at CSI.


\begin{thebibliography}{99}

\bibitem{KCNature} E. Kim and M. H. W. Chan, Nature,
{\bf 427}, 225 (2004); 

\bibitem{KCScience} E. Kim and M. H. W. Chan, Science {\bf 305}, 1941 (2004).

\bibitem{andreev69}
A. F. Andreev and I. M. Lifshitz, Sov. Phys. JETP {\bf 29}, 1107 (1969).

\bibitem{chester}
G. V. Chester, Phys. Rev. A {\bf 2}, 256 (1970).

\bibitem{Dai05}
X. Dai, M. Ma, and F.-C. Zhang, Phys. Rev. B {\bf 72}, 132504 (2005).

\bibitem{anderson}
P. W. Anderson, W. F. Brinkman and D. A. Huse, Science {\bf 310}, 1164 (2005).

\bibitem{Dorsey06} A. T. Dorsey, P. M. Goldbart and J. Toner, Phys. Rev. Lett. {\bf 96}, 055301 (2006).

\bibitem{Rittner} A. S. Rittner and J. D. Reppy, cond-mat/0604528.

\bibitem{kubota} M. Kubota {\it et al.}, unpublished (2006); 
 
\bibitem{shirahama} K. Shirahama {\it et al.}, unpublished (2006).

 
\bibitem{Bi05} J. Day, T. Herman, and J. Beamish,
Phys. Rev. Lett. {\bf 95}, 035301 (2005)

\bibitem{Bi06}
J. Day and J. Beamish,
Phys. Rev. Lett. {\bf 96}, 105304 (2006).

\bibitem{fraass}
B. A. Fraass, P. R. Granfors and R. O. Simmons, Phys. Rev. B {\bf
39}, 124 (1989).

\bibitem{meisel}
M. W. Meisel, Physica B {\bf 178}, 121 (1992).

\bibitem{theorem} N. Prokof'ev and B. Svistunov,
         Phys. Rev. Lett. {\bf 94}, 155302 (2005).

\bibitem{bernu} D. M. Ceperley and B. Bernu, Phys.
       Rev. Lett. {\bf 93}, 155303 (2004); in private communication,
the authors acknowledged that their result for $\Delta_{\rm I}$ may need to be revised.

\bibitem{superglass}
M. Boninsegni, N. Prokof'ev and B. Svistunov, Phys. Rev. Lett. {\bf 96}, 105301 (2006).

\bibitem{clark} B. K. Clark and D. M. Ceperley, Phys. Rev. Lett. {\bf 96}, 105302 (2006).

\bibitem{pederiva97}
F. Pederiva, G. V. Chester, S. Fantoni and L. Reatto, Phys. Rev. B
{\bf 56}, 5909 (1997).

\bibitem{pederiva99}
B. Chaudhuri, F. Pederiva and G. V. Chester, Phys. Rev. B {\bf 60}, 3271 (1999).

\bibitem{galli} D. E. Galli and L. Reatto, Phys. Rev. Lett. {\bf 96}, 165301 (2006).

\bibitem{worm} M. Boninsegni, N. Prokof'ev, and B. Svistunov,
Phys. Rev. Lett. {\bf 96}, 070601 (2006).

\bibitem{aziz}
R. A. Aziz, V. P. S. Nain, S. Carley, W. L. Taylor and G. T.
McConville, J. Chem. Phys. {\bf 70}, 4330 (1979).

\bibitem{TOLYAKNOWS} J. F. Jarvis, D. Ramm, and H. Meyer,
                     Phys. Rev. {\bf 170}, 320 (1968).


\bibitem{eshelby56}
J.D. Eshelby, in {\it Solid State Physics}, edited by F. Seitz and D. Turnbull, vol. {\bf 3}, p. 79 (Academic Press, New York (1956))

\bibitem{andreev76} A. F. Andreev, Sov.Phys.-JETP {\bf 41}, 1170 (1976);


\bibitem{richards75} M. G. Richards, J. H. Smith, P. S. Tofts, and W. J. Mullin,
Phys. Rev. Lett. {\bf 34}, 1545 (1975).

\bibitem{mullin75}
(W. J. Mullin, R. A. Guyer and H. A. Goldberg, Phys. Rev. Lett. {\bf 35}, 1007 (1975). 


\bibitem{khomskii01}
D.I. Khomskii, K.I. Kugel, Europhys. Lett. {\bf 55}, No. 2, 208-213 (2001);


\bibitem{burovski} E. Burovski, E. Kozik, A. Kuklov, N. Prokof'ev, and B. Svistunov
Phys. Rev. Lett. {\bf 94}, 165301 (2005).



\end{thebibliography}
\end{document}